# Normal Doppler frequency shift in negative refractive-index systems

Xiao Lin[1,*] and Baile Zhang[1,2,*]

[1]*Division of Physics and Applied Physics, School of Physical and Mathematical Sciences, Nanyang Technological University, Singapore.*

[2]*Centre for Disruptive Photonic Technologies, Nanyang Technological University, Singapore.*

[*]*Corresponding author. E-mail: xiaolinbnwj@ntu.edu.sg (X. Lin); blzhang@ntu.edu.sg (B. Zhang)*

**Abstract:** Besides the well-known negative refraction, a negative refractive-index material can exhibit another two hallmark features, which are the inverse Doppler effect and backward Cherenkov radiation. The former is known as the motion-induced frequency shift that is contrary to the normal Doppler effect, and the latter refers to the Cherenkov radiation whose cone direction is opposite to the source's motion. Here we combine these two features and discuss the Doppler effect inside the backward Cherenkov cone. We reveal that the Doppler effect is not always inversed but can be normal in negative refractive-index systems. A previously un-reported phenomenon of normal Doppler frequency shift is proposed in a regime inside the backward Cherenkov cone, when the source's velocity is two times faster than the phase velocity of light. A realistic metal-insulator-metal structure, which supports metal plasmons with an effective negative refractive index, is adopted to demonstrate the potential realization of this phenomenon.

# 1. Introduction

In his seminal work of negative refractive-index materials in 1968 [1], the Russian physicist Victor G. Veselago (1929-2018) proposed three exotic phenomena, namely, the negative refraction, the inverse Doppler effect, and the backward Cherenkov radiation (characterized by a backward Cherenkov cone opposite to the source's motion). While the absolute existence of negative refraction has caused substantial debates in past decades [2-5], the phenomena of inverse Doppler effect [6-16] and backward Cherenkov radiation [17-20] in negative refractive-index materials have never been questioned. As a result, it has been widely considered that the Doppler effect in a negative refractive-index material is always inversed, i.e., the Doppler frequency shift in such a material is negative (positive) when the source approaches (leaves) the observer. Moreover, the connection between the inverse Doppler effect and the backward Cherenkov cone has never been discussed, although both phenomena are caused by source's motion in negative refractive-index materials. Here we show that the Doppler effect in negative refractive-index materials is not always reversed, but can be normal in a regime inside the backward Cherenkov cone, when the source's velocity $v$ is two times faster than the phase velocity $v_\text{p}$ of light, i.e., $v > 2|v_\text{p}|$. Because of the superlight velocity of source, we denote this phenomenon as the superlight normal Doppler effect in negative refractive-index systems.

It is V. L. Ginzburg and I. M. Frank who noticed in 1947 that in positive refractive-index materials the Doppler effect near the Cherenkov cone exhibits anomalous properties [21-26]. They showed that when crossing the Cherenkov cone, the frequency of emitted photons will transit from positive to negative [22,23]. The criterion for this phenomenon is that the source's velocity needs to be faster than the phase velocity of light, which is the same as the Cherenkov threshold [22,23]. In other words, such an anomalous Doppler frequency shift is intrinsically connected with the Cherenkov cone. However, their discussion was limited to positive refractive-index systems.

Here we extend Ginzburg and Frank's theory of superlight Doppler effect [22,23] into negative refractive-index systems, and discuss the possibility of constructing a normal Doppler frequency shift in such systems. As a concrete example, the highly squeezed polaritons with their effective refractive index being negative, such as the negative refractive-index metal plasmons in a metal-insulator-metal structure (where the group velocity and phase velocity are antiparallel) [27,28], are adopted as a potential platform to demonstrate this normal Doppler frequency shift in negative refractive-index systems.

## 2. Results and Discussion

To facilitate the discussion of Doppler effects in the system with a negative refractive-index $n(\omega)$, we start with their analytical derivation. Without loss of generality, we consider a radiation source that moves with a velocity $\bar{v} = +\hat{z}v$ and has a natural positive angular frequency $\omega_0$ in the moving source frame. With the application of plane wave expansion [20,29], we have $\begin{bmatrix} \bar{k} \\ \omega/c \end{bmatrix} = \begin{bmatrix} \bar{\bar{\alpha}} & +\gamma\bar{\beta} \\ +\gamma\bar{\beta} & \gamma \end{bmatrix} \begin{bmatrix} \bar{k}' \\ \omega'/c \end{bmatrix}$ from the Lorentz transformation [30]; $\bar{k} = \hat{x}k_x + \hat{y}k_y + \hat{z}k_z$ ($\bar{k}'$) and $\omega$ ($\omega' = \omega_0$) are the wavevector and the frequency in the laboratory frame (the moving source frame), respectively; $\bar{\beta} = \bar{v}/c$; $\gamma = (1 - \beta^2)^{-1/2}$ is the Lorentz factor; $\bar{\bar{\alpha}} = \bar{\bar{I}} + (\gamma - 1)\frac{\bar{\beta}\bar{\beta}}{\beta^2}$; $\bar{\bar{I}}$ is the unity dyad. Then it is straightforward to obtain $k_z = \frac{\omega - \omega_0/\gamma}{v}$. Moreover, the relation between $k_z$ and $\bar{k}$ indicates that

$$k_z = |n(\omega)|\frac{\omega}{c} \cdot (-cos\theta) \tag{1}$$

where the radiation angle $\theta$ [Figure 1] is the angle between $\bar{v}$ and $\bar{S}$. Here the Poynting vector $\bar{S}$ is antiparallel to the wavevector $\bar{k}$ in negative refractive-index systems, and $|\bar{k}| = |n(\omega)|\frac{\omega}{c}$. With the combination of the two equations for $k_z$, we have

$$\omega - \omega_0/\gamma = -|n(\omega)|\frac{v}{c}\omega cos\theta \tag{2}$$

If $|n(\omega)|\frac{v}{c}cos\theta \neq -1$ and $\omega_0 \neq 0$ (e.g., the source is a moving dipole), the Doppler effect is governed by

$$\omega = \frac{\omega_0/\gamma}{1+|n(\omega)|\frac{v}{c}cos\theta} \tag{3}$$

The appearance of $\gamma$ in equation (3) is due to the time dilation or the relativistic effect [30].

If $|n(\omega)|\frac{v}{c}\cos\theta = -1$, equation (3) is satisfied only if $\omega_0 = 0$ (namely the source is a moving charged particle, instead of a moving dipole for the Doppler effect). This corresponds to the backward Cherenkov radiation [1,17-20] and is featured by a backward Cherenkov cone, where its opening angle $\theta_{CR}$ satisfies $v\cos\theta_{CR} = c/n$.

Although the Doppler effect and backward Cherenkov radiation in negative refractive-index systems are two different physical phenomena, they have a strong connection via the backward Cherenkov cone. In a nutshell, the backward Cherenkov cone divides the K-space of the Doppler effect into two parts, each of which has different properties. This can be understood as follows.

If $\theta < \theta_{CR}$, equation (3) is valid only for $\omega > 0$, since $|n(\omega)|\frac{v}{c}\cos\theta > -1$; this corresponds to the conventional inverse Doppler effect in negative refractive-index systems as proposed by Veselago [1]. Namely, the Doppler frequency shift outside the backward Cherenkov cone is always inversed in negative refractive-index systems [Figure 1]. To be specific, the observer will receive a frequency higher (lower) than the emitted frequency during the recession (approach), i.e., $\Delta\omega > 0$ at $\theta > 90^o$ ($\Delta\omega < 0$ at $\theta < 90^o$) [Figure 1], where $\Delta\omega = |\omega| - \omega_0/\gamma$ is the Doppler frequency shift.

In contrast, if $\theta > \theta_{CR}$, equation (3) is valid only for $\omega < 0$, since $|n(\omega)|\frac{v}{c}\cos\theta < -1$. The negative $\omega$ was firstly revealed by Ginzburg and Frank's theory of superlight Doppler effects in positive refractive-index systems [21-23]. There are many other exotic phenomena related to negative frequencies [31-35], such as the fiber-optical analog of the event horizon [34,35]. Therefore, if $|n(\omega)|\frac{v}{c}\cos\theta < -1$, equation (3) corresponds to the superlight Doppler effects in negative refractive-index systems. By following Ginzburg and Frank's terminology [22,23], we denote the cone in the K-space satisfying $|n(\omega)|\frac{v}{c}\cos\theta = -1$ in negative refractive-index systems as the backward Cherenkov cone [Figure 1]. If

the frequency dispersion is neglected, i.e., $n(\omega)$ is a constant, $\omega$ diverges and transits from positive infinity to negative infinity when crossing the backward Cherenkov cone.

Inside the backward Cherenkov cone, the superlight Doppler effects in negative refractive-index systems can be divided into two types. If $-2 < |n(\omega)|\frac{v}{c}\cos\theta < -1$ in equation (3), the Doppler frequency shift inside the backward Cherenkov cone is still inversed, i.e., $\Delta\omega > 0$ at $\theta > 90^o$ [Figure 1]. This is denoted as the superlight *inverse* Doppler effect in negative refractive-index systems.

If $|n(\omega)|\frac{v}{c}\cos\theta < -2$ in equation (3), we find an un-reported Doppler phenomenon inside the backward Cherenkov cone. To be specific, the Doppler frequency shift in negative refractive-index systems becomes normal in a regime inside the backward Cherenkov cone, i.e., $\Delta\omega < 0$ at $\theta > 90^o$ [Figure 1]. This corresponds to the superlight *normal* Doppler effect in negative refractive-index systems. In addition, we show more analyses of the K-space representation of the Doppler effect at different values of $v$ in Supporting Information Figure S3.

To get a straightforward understanding of the normal Doppler frequency shift in negative refractive-index systems, we illustrate the Doppler effect in time domain in Figure 2. Figure 2a shows multiple wave fronts equally-distributed in phase and with different radii around a stationary source. Note that in negative refractive-index systems, the circular wave fronts propagate inward toward the source (while the wave energy propagates outward instead). So an observer will firstly receive wave fronts with smaller radii before wave fronts with larger radii arrive. When the source moves with a velocity of $v < 2|v_\text{p}|$, as in Figure 2b,c, the wave fronts bunch together at $\theta = 180^o$ or in the backward direction, where $v_\text{p} = c/n$ is the phase velocity of light. This leads to an inversed Doppler frequency shift, i.e., $\Delta\omega > 0$ at $\theta = 180^o$ in negative refractive-index systems. In contrary, when the source moves with a velocity of $v > 2|v_\text{p}|$ in Figure 2d, the wave fronts spread out at $\theta = 180^o$, giving rise to the normal Doppler frequency shift, i.e., $\Delta\omega < 0$ at $\theta = 180^o$ in negative refractive index systems. It is worthy to note that when the source moves at a subluminal velocity ($v < |v_\text{p}|$), as in Figure 2b, an observer at $\theta = 180^o$ will firstly receive wave fronts with smaller radii (which corresponds to $\omega > 0$ in equations (3)), the same as Figure 2a. In contrast, if the source moves at a superlight velocity ($v > |v_\text{p}|$), as in Figure 2c,d, an observer at $\theta = 180°$ will firstly receive wave fronts with larger radii (which corresponds to $\omega < 0$ in equations (3)).

To facilitate the potential observation of the revealed normal Doppler frequency shift in Figures 1 and 2, a realistic negative refractive-index system is needed. Ever since the advent of metamaterials, the negative refractive-index systems have attracted enormous attentions [2-6,8,9,19,36]. In Veselago's seminal work [1], a negative refractive-index system refers to a material with simultaneously negative permittivity and permeability. Such a double negative material does not naturally exist but can be artificially constructed, such as through the design of metamaterials [36] and photonic crystals [8,37], but its realization is generally challenging. In addition to double negative materials, the effective negative refractive index for eigenmodes in some plasmonic or waveguide systems have also been reported, if the directions of the phase and group velocities for these eigenmode are anti-parallel (this circumvents the requirement for the permittivity and permeability to be simultaneously negative). For example, the effective negative refractive index has been realized for highly squeezed polaritons, e.g., metal plasmons in a metal-insulator-metal structure as shown in Figure 3 [27,28] and phonon polaritons in a thin slab of hexagonal boron nitride [38-41].

Before we proceed, it shall be emphasized that the analytical derivation for the Doppler effect in equations (1-3) is also applicable to plasmonic or waveguide systems, if they support eigenmodes propagating in a plane parallel to the source's trajectory (e.g., waves guided in the $x$-$z$ plane but confined along the $y$ direction ) and if these eigenmodes have an effective negative refractive index $n_\text{eff}(\omega)$. In such systems, the Poynting vector $\bar{S}$ becomes antiparallel to the in-plane wavevector $\bar{k}_\| = \hat{x}k_x + \hat{z}k_z$ (the component parallel to the $x$-$z$ plane) of these eigenmodes and the effective in-plane refractive index is defined as $n_\text{eff}(\omega) = k_\|/(\omega/c)$. For plasmonic systems such as metal plasmons in the metal-insulator-metal structure [27,28], $\bar{k}_\| = \bar{k}_\text{spp}$, where $\bar{k}_\text{spp}$ is the in-plane wavevector of plasmonic eigenmodes.

Accordingly, for plasmonic or waveguide systems, $n(\omega)$ in equations (1-3) shall be replaced with $n_{\text{eff}}(\omega)$, and $\theta$ becomes the angle between $\bar{S}$ (or $-\bar{k}_{\parallel}$) and $\bar{v}$; see the inset of Figure 4 for example.

From equation (3) and the above analysis, the deterministic factor for the Doppler effect is the effective refractive index, instead of the effective permittivity or permeability. Therefore, it is feasible to adopt some plasmonic or waveguide systems, which support the propagation of eigenmodes with an effective negative refractive index, to demonstrate the superlight normal Doppler effect in Figures 1 and 2. As a typical example, the negative refractive-index metal plasmons are adopted in the following discussions.

Figure 3 shows the dispersion of metal plasmons in a metal-insulator-metal structure. For metal plasmons in this plot, we pre-set the value of their group velocity $v_g = \frac{\partial \omega}{\partial k_{\text{spp}}}$ to be positive, i.e., $v_g > 0$, to faciliate the clear definition of their effective refractive index. Correspondingly, the phase velocity for metal plasmons is $v_p = \frac{\omega}{k_{\text{spp}}}$. If the directions of the phase and group velocities are anti-parallel in the $x$-$z$ plane, we have $v_g \cdot v_p < 0$ and thus $v_p < 0$. Since $v_p < 0$, we have $\text{Re}(k_{\text{spp}}) < 0$ if $\omega > 0$ in Figure 3. This way, the effective refractive index of metal plasmons can be directly defined as $n_{\text{eff}}(\omega) = k_{\text{spp}}/(\omega/c)$. From the principle of causality, we also have $n_{\text{eff}}(-\omega) = n_{\text{eff}}^*(\omega)$ [30]. In Figure 3, the experimental data of permittivity of silver [42] is adopted. The insulator has a refractive index of 3 (e.g., boron phosphide [43]) and a thickness of $d = 0.01\lambda_p$, where $\lambda_p = 300$ nm. Figure 3 shows that $n_{\text{eff}}(\omega)$ for metal plasmons is negative within the range of $[0.6\ 0.9]\omega_p$, where $\omega_p = 2\pi/\lambda_p$. Moreover, it is noted that for negative refractive-index systems, the absolute value of refractive index generally decreases with the frequency, such as metal plasmons in Figure 3.

Figure 4 shows the possible realization of the normal Doppler frequency shift for the negative refractive-index metal plasmons. Below we consider the working frequencies only in the above range and let the dipole move parallel to the interfaces of metal-insulator-metal; see the structural setup in the inset of Figure 3. Note that the vertical distance between the moving dipole and the interface of metal-insulator-metal has a trivial influence on the Doppler frequency shift in Figure 4, although it may affect the field distribution of excited metal plasmons. Two cases, i.e., $v = 0.08c$ and $v = 0.8c$, are studied in Figure 4. It shall be emphasized that the frequency dispersion of negative refractive-index systems is neglected for conceptual demonstration only in Figures 1 and 2 and also in Veselago's seminal work [1], but is unavoidable in reality [44] such as the realistic system in Figures 3 and 4. When considering the dispersion, we note that the Doppler effect in negative refractive-index systems will be highly dependent on the dispersion. For example, due to the specific dispersion of metal plamsons (where $|n_{\text{eff}}(\omega)|$ decreases with frequency), there are two unique phenomena related to the superlight normal Doppler effect in negative refractive-index systems.

First, the superlight normal and conventional inverse Doppler effects for negative refractive-index metal plasmons may simultaneously show up at the same values of $\theta$ in Figure 4. For example, when $v = 0.08c$ in Figure 4, we simultaneously have $\omega = 0.96\omega_0/\gamma$ (which corresponds to the superlight normal Doppler effect in negative refractive-index systems) and $\omega = 1.21\omega_0/\gamma$ (conventional inverse Doppler effect in negative refractive-index systems) at $\theta = 150^o$. This phenomenon can be explained as follows. When considering the dispersion in negative refractive-index systems, the backward Cherenkov cone in the interested frequency range can still be well defined as the cone in the K-space that has the minimum value of $\theta$ satisfying $|n_{\text{eff}}(\omega)|\frac{v}{c}\cos\theta = -1$. According to the definition of the backward Cherenkov cone in dispersion-less and dispersive systems, the superlight Doppler effects in negative refractive-index systems always appear inside the backward Cherenkov cone. In contrast, the conventional inverse Doppler effect in negative refractive-index systems appears only outside the backward Cherenkov cone when the dispersion is neglected as in Figure 1, but may appear inside the backward Cherenkov cone when the frequency dispersion is considered as in Figure 4. In other words, when considering the frequency dispersion, the superlight Doppler effects and the conventional inverse Doppler effect in negative refractive-index systems may be partially overlapped in the K-space inside the backward Cherenkov cone, as shown in Figure 4.

Second, the superlight normal Doppler effect might have even a lower velocity threshold than the superlight inverse Doppler effect. In absence of the frequency dispersion, one shall always expect that, the appearance of superlight normal Doppler effect (which requires $v > 2c/|n|$ from equation (3)) always has a larger threshold of $v$ than the superlight inverse Doppler effect (which only needs $v > c/|n|$ from equation (3)); see more analysis in Supporting Information Figure S3. As a result, inside the backward Cherenkov cone in Figure 1, the regime of superlight normal Doppler effect should be always wrapped around by the regime of superlight inverse Dopper effect. However, this rule is not applicable when considering the realistic dispersion of negative refractive-index systems. As a representative example, when $v = 0.08c$ in Figure 4, for superlight Doppler effects of negative refractive-index metal plasmons, there is only the superlight normal Doppler effect, without the appearance of the superlight inverse Doppler effect. To faciliate the explanation of this example, we choose two frequencies of $\omega_{\text{low}}$ and $\omega_{\text{high}}$ in the neighborhood of $\omega_0/\gamma$, and let $|\omega_{\text{low}}| < \omega_0/\gamma < |\omega_{\text{high}}|$. For metal plasmons with $|n_{\text{eff}}(\omega)|$ decreasing with frequency, one has $|n_{\text{eff}}(\omega_{\text{low}})| > |n_{\text{eff}}(\omega_{\text{high}})|$, which may lead to $c/|n_{\text{eff}}(\omega_{\text{high}})| > v > 2c/|n_{\text{eff}}(\omega_{\text{low}})|$ [e.g., when $|n_{\text{eff}}(\omega_{\text{low}})| > 2|n_{\text{eff}}(\omega_{\text{high}})|$]. If $c/|n_{\text{eff}}(\omega_{\text{high}})| > v > 2c/|n_{\text{eff}}(\omega_{\text{low}})|$, we can have the situation that the emergence condition of superlight normal Doppler effect (i.e., $v > 2c/|n_{\text{eff}}(\omega_{\text{low}})|$) is fulfilled, while the emergence condition of superlight inverse Doppler effect ($v > c/|n_{\text{eff}}(\omega_{\text{high}})|$) is failed. In other words, when considering the realistic dispersion, the appearance of superlight normal Doppler effect does not always need to have a larger threshold of $v$ than the superlight inverse Doppler effect in negative refractive-index systems. By using this fact, we also show in Supporting Information Figure 2 the possibility to spatially separate the superlight normal Doppler effect from the other Doppler effects in negative refractive-index systems, which may facilitate its experimental observation.

## 3. Conclusion

In conclusion, we have revealed the possibility to create the normal Doppler frequency shift in negative refractive-index systems, by finding an un-reported superlight normal Doppler effect in a regim inside the backward Cherenkov cone. This way, our work further develops Veselago's theory of inverse Doppler effect in negative refractive-index systems and Ginzburg & Frank's theory of superlight Doppler effects in positive refractive-index systems. In addition to the isotropic (for both negative and positive refractive-index) systems, there are many other anisotropic systems, such as systems supporting hyperbolic eigenmodes (e.g., hyperbolic metamaterials) and photonic or plasmonic systems supporting nonreciprocal eigenmodes. The continuing exploration of Doppler effects at different velocities of the moving source in these intriguing systems [45,46] is highly wanted.

**Supporting Information**
Supporting Information is available from the Wiley Online Library or from the author.

**Acknowledgements**
This work was sponsored by Nanyang Technological University for NAP Start-Up Grant and the Singapore Ministry of Education (Grant No. MOE2015-T2-1-070, MOE2016-T3-1-006 and Tier 1 RG174/16 (S)).

**Conflict of Interest**
The authors declare no conflict of interest.

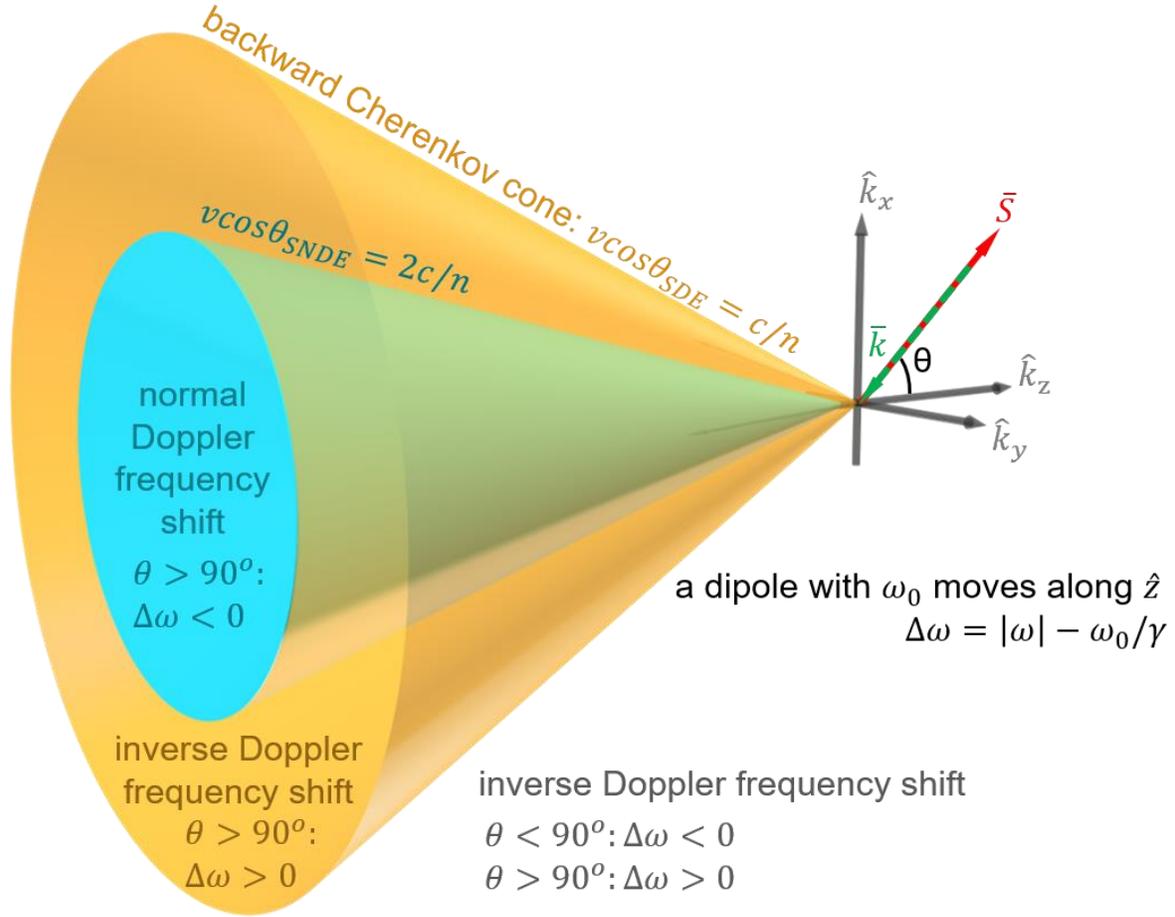

**Figure 1.** K-space representation of the normal Doppler frequency shift in negative refractive-index systems. The normal Doppler frequency shift can appear in a regime (blue region) inside the backward Cherenkov cone. A source or dipole moves with a velocity $\bar{v} = \hat{z}v$. In the moving source frame, the source has a natural frequency $\omega_0$ ($\omega_0 \neq 0$). In the laboratory frame, the received radiation fields have a frequency $\omega$, a wavevector $\bar{k}$, and the corresponding Poynting vector (i.e, the power flow density) $\bar{S}$, where $\bar{S}$ is antiparallel to $\bar{k}$; $\theta$ is the angle between $\bar{S}$ and $\bar{v}$ (or the $\hat{k}_z$ axis); $\gamma$ is the Lorentz factor. The angles of $\theta_{SDE}$ and $\theta_{SNDE}$ are the opening angles of the cone in which the superlight Doppler effects (blue and orange regions) occur and of the cone in which the superlight normal Doppler effect occurs, respectively. Note that $\theta_{SDE} = \theta_{CR}$, where $\theta_{CR}$ is the opening angle of the backward Cherenkov cone created by a moving charged particle (instead of a dipole here), since both $\theta_{SDE}$ and $\theta_{CR}$ satisfy the condition of $v \cos\theta = c/n$. For the conceptual illustration, we let $v > 2c/|n|$ here and set $n$ to be a negative constant here and in Figure 2.

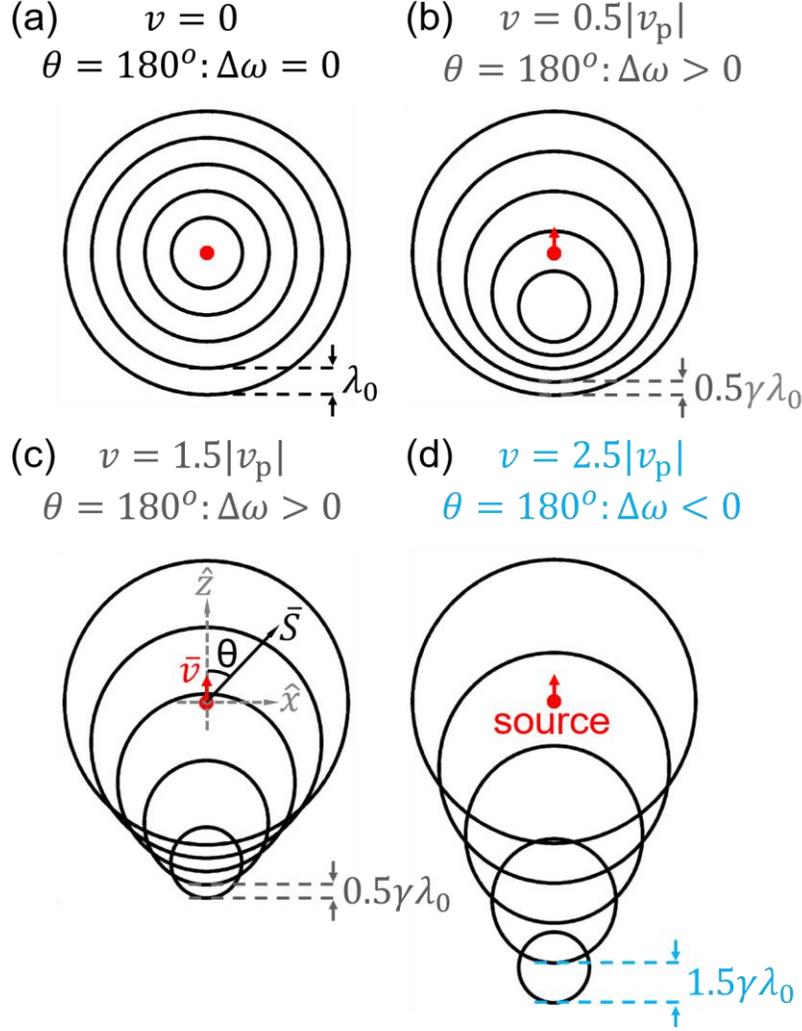

**Figure 2.** Real-space illustration of the normal Doppler frequency shift in a negative refractive-index system. The multiple wave fronts, illustrated by circular lines, are equally-distributed in time. $\theta$ is the angle between $\bar{S}$ and $\bar{v}$. **a** If $v = 0$, there is no Doppler frequency shift, i.e., $\Delta\omega = 0$, where $\Delta\omega = |\omega| - \omega_0/\gamma$. **b, c** If $v < 2|v_p|$, the wave fronts bunch together at $\theta = 180°$, where $v_p = c/n$ is the phase velocity of light. This leads to an inverse Doppler frequency shift, i.e., $\Delta\omega > 0$ at $\theta = 180°$. **d** If $v > 2|v_p|$, the wave fronts spread out at $\theta = 180°$, leading to a normal Doppler frequency shift, i.e., $\Delta\omega < 0$ at $\theta = 180°$. The Lorentz factor $\gamma$ is different in each panel. Here we set $n < -2.5$ for illustration.

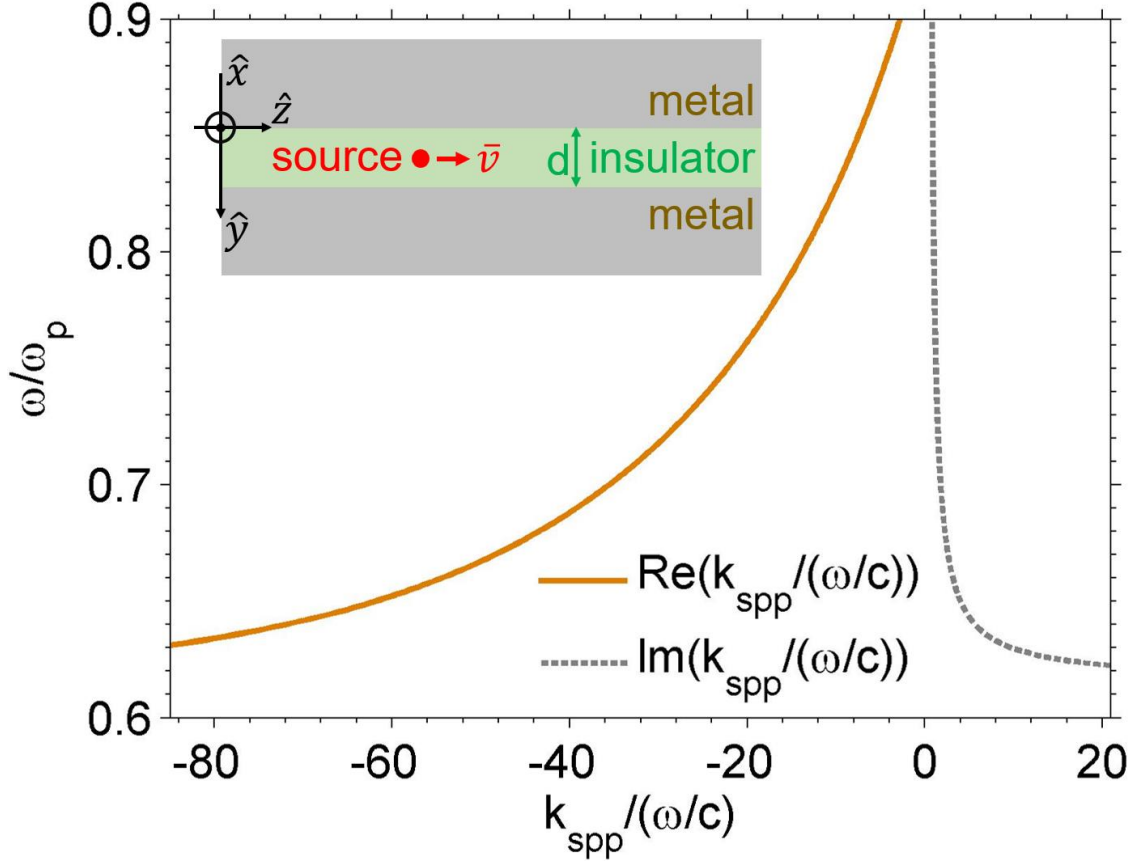

**Figure 3.** Dispersion of negative refractive-index metal plasmons in a metal-insulator-metal structure. For metal plasmons, $\bar{k}_{spp} = \hat{x}k_x + \hat{z}k_z$ is the component of wavevector parallel to the $x$-$z$ plane or the interface. This figure is plotted by setting the group velocity of metal plasmons to be positive, i.e., $\frac{\partial \omega}{\partial \text{Re}(k_{spp})} > 0$. This way, the effective refractive index of metal plasmons can be defined as $n_{eff}(\omega) = k_{spp}/(\omega/c)$. The real part of $n_{eff}(\omega)$ is negative and its absolute value *decreases* with frequency, when the frequency is within the range of $[0.6\ 0.9]\omega_p$, where $\omega_p = 2\pi/\lambda_p$ and $\lambda_p = 300$ nm. For the negative refractive-index metal plasmons, we have $\text{Re}(k_{spp}) \cdot \text{Im}(k_{spp}) < 0$; in other words, $k_{spp}$ is in the second or fourth quadrant of the complex $k_{spp}$ plane. The metal-insulator-metal structure is shown in the inset. The experimental data of permittivity of silver [42] is adopted. The insulator has a refractive index of 3 (e.g., boron phosphide [43]) and a thickness of $d = 0.01\lambda_p$.

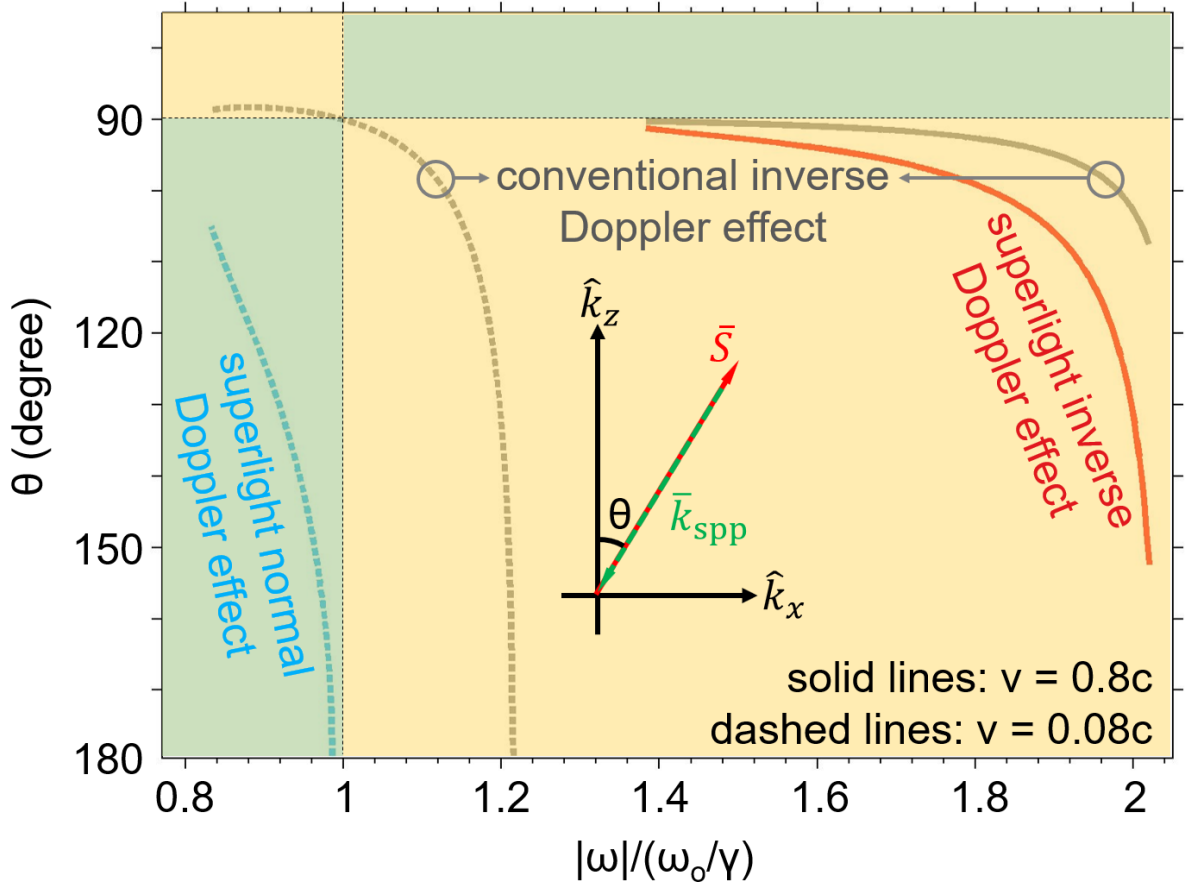

**Figure 4.** Normal Doppler frequency shift of negative refractive-index metal plasmons in a metal-insulator-metal structure. The Doppler frequency shift is normal in the blue regions (namely $\Delta\omega < 0$ at $\theta > 90^o$ and $\Delta\omega > 0$ at $\theta < 90^o$), but is inversed in the yellow regions ($\Delta\omega > 0$ at $\theta > 90^o$ and $\Delta\omega < 0$ at $\theta < 90^o$). A dipole moves parallel to the interface of the meta-insulator-metal structure and excites metal plasmons. The propagating angle $\theta$ for excited metal plasmons in the $x$-$z$ plane is illustrated in the inset; the corresponding structural setup is shown in the inset of Figure 3. The dipole moves with a velocity $\bar{v} = \hat{z}v$ and has $\omega_0 = 0.75\omega_p$, where $\omega_p = 2\pi/\lambda_p$ and $\lambda_p = 300$ nm. The effective refractive index for metal plasmons is negative, e.g., $n(0.75\omega_p) = -22$, and dcreases with frequency in the studied frequecy range; see Figure 3. The normal Doppler frequency shift of metal plasmons can appear if $v = 0.08c$, i.e., the superlight normal Doppler effect with $\Delta\omega < 0$ at $\theta > 90^o$.